\documentclass[final,5p,times,twocolumn]{elsarticle}
\usepackage{graphicx,dcolumn,bm,amssymb,amsmath,latexsym,amsfonts,amsthm}

\journal{Physics Letters B}

\begin{document}
%%%%%%%%%%%%%%%%%%%%%%%%%%%%%%%%%%%%%
%defining some commands
\newcommand{\nonu}{\nonumber}
\newcommand{\sm}{\small}
\newcommand{\noi}{\noindent}
\newcommand{\npg}{\newpage}
\newcommand{\nl}{\newline}
\newcommand{\bp}{\begin{picture}}
\newcommand{\ep}{\end{picture}}
\newcommand{\bc}{\begin{center}}
\newcommand{\ec}{\end{center}}
\newcommand{\be}{\begin{equation}}
\newcommand{\ee}{\end{equation}}
\newcommand{\beal}{\begin{align}}
\newcommand{\eeal}{\end{align}}
\newcommand{\bea}{\begin{eqnarray}}
\newcommand{\eea}{\end{eqnarray}}
\newcommand{\bnabla}{\mbox{\boldmath $\nabla$}}
\newcommand{\univec}{\textbf{a}}
\newcommand{\VectorA}{\textbf{A}}
\newcommand{\Pint}
%%%%%%%%%%%%%%%%%%%%%%%%%%%%%%%%%%%%

\begin{frontmatter}

\title{Corotating two-body system of identical Kerr sources}

\author[one]{I. Cabrera-Munguia\corref{cor1}}
\cortext[cor1]{Corresponding author.}
\ead{icabreramunguia@gmail.com}
\author[two]{V. E. Ceron}
\ead{vceron@uaeh.edu.mx}
\author[two]{L. A. L\'opez}
\ead{lalopez@uaeh.edu.mx}
\author[two]{Omar Pedraza}
\ead{omarp@uaeh.edu.mx}
\address[one]{Departamento de F\'isica y Matem\'aticas, Universidad Aut\'onoma de Ciudad Ju\'arez, 32310 Ciudad Ju\'arez, Chihuahua, M\'exico}
\address[two]{\'Area Acad\'emica de Matem\'aticas y F\'isica, UAEH, carretera Pachuca-Tulancingo km 4.5, C.P. 42184, Pachuca, Hidalgo, M\'exico}

\begin{abstract}
A binary system of equal corotating Kerr sources is studied after deriving the corresponding $3$-parametric asymptotically flat exact solution. Both sources are apart from each other by means of a massless strut (conical singularity). In the context of black holes, the analytical functional form of each horizon $\sigma$ is expressed in terms of arbitrary Komar physical parameters: mass $M$, angular momentum $J$ (with parallel spin), and the coordinate distance $R$. Later on, all the thermodynamical properties related to the horizon are depicted by concise formulae. Finally, the extreme limit case is obtained as a $2$-parametric subclass of Kinnersley-Chitre metric.
\end{abstract}


\end{frontmatter}

\section{Introduction}
\vspace{-0.3cm}In the context of exact solutions in general relativity, dynamical scenarios involving two rotating black hole (BH) sources turn out to be quite difficult to understand, since there are complicated issues regarding physical effects produced by multipolar interactions between sources, as well as the construction of the exact solution itself. Perhaps this is the main reason which has been leading us to study frequently stationary axisymmetric spacetimes concerning Kerr-type sources after the double-Kerr-NUT (DKN) came to light in 1980 \cite{KramerNeugebauer}. Especially, in the search of equilibrium configurations (without a supporting strut), where the notion of treating with BH sources is ruined due to the presence of ring singularities off the axis \cite{Hennig,DH} which are associated to its negative mass \cite{MRS1}. The latter situation endorses the idea of making efforts for searching new binary models describing rotating Kerr (or Kerr-Newman) sources with a massless strut in between (conical singularity \cite{BachW,Israel}), and provide novel evidence on their physical properties. Unfortunately such task remains complicated to perform since the axis conditions in the most general case has not yet been solved until the present day.

After settling the appropriate Riemann-Hilbert problem, Varzugin \cite{Varzugin} provided various dynamical and thermodynamical aspects related to rotating BH sources. In particular, for identical BHs with opposite spin (a counterrotating system), he showed that the interaction force related to the strut, seems to be equal compared with two identical Schwarzschild BHs \cite{BachW}. Even more, he derived an explicit formula for the angular velocity at the horizon which led straightforwardly to the corresponding horizon $\sigma$ in terms of Komar physical parameters \cite{Komar}: the mass $M$ and angular momentum $J$, and the coordinate distance $R$ between the centers of the horizons, namely
\be \sigma= \sqrt{M^{2}-\frac{J^{2}}{M^{2}}\left(\frac{R-2M}{R+2M}\right)}.  \label{counter}\ee

On the other hand, regarding the corotating sector (sources with aligned spin), Costa \emph{et al.} \cite{Costa} improved Varzugin's work offering more physical and thermodynamical properties. Nevertheless, those authors delivered only a numerical study of the solution, since they never obtained a similar expression for the horizon $\sigma$ like the aforementioned Eq.\ (\ref{counter}) for counterrotating two-body systems. It should be pointed out, that the knowledge of $\sigma$ as a function of Komar parameters defines in a more transparent way the whole structure and its geometrical (thermodynamical) properties of the spacetime, but it is not trivial to perform such a task.

The present paper aims at the construction of a $3$-parametric physical model describing a two-body system composed by identical corotating Kerr sources apart by a massless strut. To reach our goal, we adopt a suitable parametrization for solving the axis conditions, and later on, to provide an explicit but nontrivial formula for the horizon half-length parameter $\sigma$ in terms of Komar physical parameters $\{R,M,J\}$. In addition, all the thermodynamical characteristics contained into the Smarr formula \cite{Smarr} are obtained. Finally, the $2$-parametric subclass of the Kinnersley-Chitre metric \cite{KCH} concerning to the extreme limit case is derived and presented in a closed analytical form by using the Perj\'es' factor structure \cite{Perjes}.

\vspace{-0.3cm}\section{Three parametric exact solution}
\vspace{-0.2cm}It is well-known that Ernst's formalism \cite{Ernst} reduces stationary axisymmetric vacuum spacetimes into a new complex equation
\be ({\cal{E}}+ \bar{\cal{E}})({\cal{E}}_{\rho \rho} + \rho^{-1}{\cal{E}}_{\rho}+{\cal{E}}_{z z} )=2({\cal{E}}_{\rho}^{2}+{\cal{E}}_{z}^{2}), \label{Ernsteq}\ee

\noi where ${\cal{E}}$ is the so-called Ernst potential. It follows that any solution of Eq.\ (\ref{Ernsteq}) permits us to derive the metric functions $f$, $\omega$, and $\gamma$ of the line element
\cite{Papapetrou}
\be ds^{2}=f^{-1}\left[e^{2\gamma}(d\rho^{2}+dz^{2})+\rho^{2}d\varphi^{2}\right]- f(dt-\omega d\varphi)^{2},
\label{Papapetrou}\ee

\noi after solving the following set of differential equations:
\bea \begin{split}  f&=  {\rm{Re}}({\cal{E}}), \\
\omega_{\rho} &= -4\rho ({\cal{E}}+ \bar{\cal{E}})^{-2}{\rm{Im}}({\cal{E}}_{z}) ,\\
\omega_{z} &= 4\rho ({\cal{E}}+ \bar{\cal{E}})^{-2}{\rm{Im}}({\cal{E}}_{\rho}) ,\\
\gamma_{\rho}&=\rho ({\cal{E}}+ \bar{\cal{E}})^{-2} \left({\cal{E}}_{\rho} \bar {\cal{E}}_{\rho} -{\cal{E}}_{z} \bar {\cal{E}}_{z}\right),\\
\gamma_{z}&=2\rho ({\cal{E}}+ \bar{\cal{E}})^{-2} \rm{Re}({\cal{E}}_{\rho}\,{\bar{\cal{E}}}_{z}).
\label{metrics}\end{split}\eea

\begin{figure}[ht]
\centering
\includegraphics[width=6.0cm,height=5.0cm]{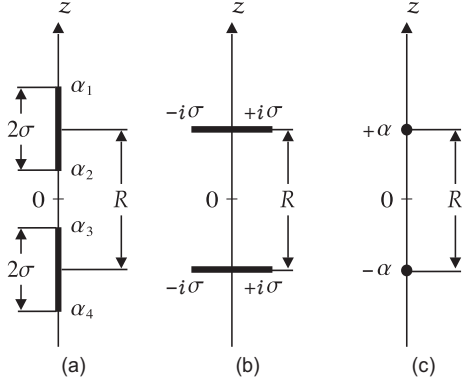}
\caption{Identical Kerr sources on the symmetry axis, with the values $\alpha_{1}= -\alpha_{4}=R/2 +\sigma$, \, $\alpha_{2}= -\alpha_{3}=R/2 -\sigma$: (a) BH configuration $\sigma^{2}>0$; (b) hyperextreme sources if $\sigma \rightarrow i \sigma$ (or $\sigma^{2}<0$ ); (c) the extreme limit case $\sigma=0$.}
\label{DKidentical}\end{figure}

In the above description the bar over a symbol refers to a complex conjugation, while a subscript $z$ or $\rho$ defines partial differentiation. Then, to solve the non-linear Eq.\ (\ref{Ernsteq}) and describe a binary system of Kerr sources, it can be used the well-known Sibgatullin's method (SM) \cite{Sibgatullin} which takes the axis data  and allows us to construct the Ernst potential ${\cal{E}}(\rho,z)$ in the entire spacetime. In this context, the extended DKN problem \cite{KramerNeugebauer} is performed directly by using the last formulas of \cite{RMJ}, with $N=2$, and after eliminating the electromagnetic field ($\Phi=0$). In fact, the full metric contains into the set $\{\alpha_{n},\beta_{j}\}$ eight algebraic parameters, where $n=\overline{1,4}$ and $j=1,2$. An asymptotically flat exact solution can be carried out settling first the \emph{axis conditions}. The extended DKN problem within the framework of SM \cite{RMJ} was constructed in such a way that the metric functions, $\omega$, and $\gamma$ automatically satisfy the conditions: $\omega(\rho=0,\alpha_{1}< z <\infty)=0$, and $\gamma(\alpha_{1}<z<\infty)=\gamma(\rho=0,-\infty<z<\alpha_{4})$; it means that an \emph{elementary flatness} is established on the upper part of the symmetry axis (see Fig.\ \ref{DKidentical}). In order to define regularity of the metric we must impose two additional conditions on the remaining parts of the symmetry axis, namely
\be \omega(\rho=0, \alpha_{2}<z< \alpha_{3})=0, \quad \omega(\rho=0, -\infty<z< \alpha_{4})=0,  \ee

\noi thereby one gets a simple representation of these axis conditions given in \cite{ICM}
\bea \begin{split} {\rm{Im}}\left[ \left|\begin{array}{ccccc}
0 &     1       &        1     &       1    &      1 \\
1 & \pm \gamma_{11}  & \pm \gamma_{12}  & \gamma_{13}& \gamma_{14}  \\
1 & \pm \gamma_{21} & \pm \gamma_{22} & \gamma_{23} & \gamma_{24} \\
0 & \kappa_{11} & \kappa_{12} & \kappa_{13} & \kappa_{14}\\
0 & \kappa_{21} & \kappa_{22} & \kappa_{23} & \kappa_{24}\\
\end{array}
\right| \right]=0,& \\
\gamma_{jn}=(\alpha_{n}-\beta_{j})^{-1} \qquad \kappa_{jn}=(\alpha_{n}-\bar{\beta}_{j})^{-1}.&
\end{split}\label{omegasregions}\eea

The first condition (with $+$ sign) eliminates the gravitomagnetic monopole (NUT charge) \cite{NUT} while the second one (with $-$ sign) ensures that the mass on the middle region does not contribute to the total ADM mass \cite{ADM}. Therefore, after solving these algebraic equations one makes sure that the total mass and total angular momentum of the system are the sum of the individual components. In this regard, the easiest solution defining a corotating binary system composed by identical sources is achieved by establishing the relation $\beta_{1} + \beta_{2}=-2M + 2iq$, and locating the sources on the symmetry axis as shown in Fig.\ \ref{DKidentical}. For this particular situation the explicit result of solving the axis conditions is expressed as
\bea \begin{split} \beta_{1,2}&= -M+i q \pm\sqrt{p+i (\delta-2M q)},\\
p&=R^{2}/4+ \delta^{2}/(R^{2}-4M^{2}+4q^{2}), \\
\delta&= \frac{2Mq(R^{2}-4M^{2}+4q^{2})}{R(R+2M)+4q^{2}}, \label{explicitsolution}\end{split}\eea

\noi where $\sigma$ defines the half-length of each rod (see Fig.\ \ref{DKidentical}), and it assumes the form
\be \sigma= \sqrt{M^{2}-q^{2}\left(1-\frac{4M^{2}(R^{2}-4M^{2}+4q^{2})}{[R(R+2M)+4q^{2}]^{2}}\right)}. \label{horizon}\ee

The Ernst potential and full metric can be worked out easily leading us to
\bea \begin{split}
{\cal{E}}&=\frac{\Lambda+\Gamma}{\Lambda-\Gamma},\qquad
f=\frac{\Lambda \bar{\Lambda}-\Gamma \bar{\Gamma}}{(\Lambda-\Gamma)(\bar{\Lambda}-\bar{\Gamma})}, \\ \omega&=\frac{2{\rm{Im}}\left[(\Lambda-\Gamma)(z\bar{\Gamma}+\bar{\mathcal{G}})\right]}{\Lambda \bar{\Lambda}-\Gamma \bar{\Gamma}},\qquad e^{2\gamma}=\frac{\Lambda \bar{\Lambda}-\Gamma \bar{\Gamma}}{256\sigma^{4}R^{4}\kappa_{o}^{2} r_{1}r_{2}r_{3}r_{4}},\\
\Lambda&=4\sigma^{2}(p_{+}p_{-}s_{+}s_{-}r_{1}r_{2}+ \bar{p}_{+}\bar{p}_{-}\bar{s}_{+}\bar{s}_{-}r_{3}r_{4})\\
&-R^{2}(\bar{p}_{+}\bar{p}_{-}s_{+}s_{-}r_{1}r_{3} + p_{+}p_{-} \bar{s}_{+}\bar{s}_{-}r_{2}r_{4} )\\
&+(R^{2}-4\sigma^{2})(\bar{p}_{+}p_{-}\bar{s}_{+}s_{-}r_{1}r_{4} + p_{+}\bar{p}_{-}s_{+}\bar{s}_{-}r_{2}r_{3}),\\
\Gamma&=-2i\sigma R\{(R-2\sigma){\rm Im}\left(p_{+}\bar{p}_{-}\right)(s_{+}s_{-}r_{1}-\bar{s}_{+}\bar{s}_{-}r_{4})\\
&+(R+2\sigma){\rm Im}\left(s_{+}\bar{s}_{-}\right)(p_{+}p_{-}r_{2}-\bar{p}_{+}\bar{p}_{-}r_{3})\},\\
\mathcal{G}&=  4\sigma^{2}\left[(R-2iq)p_{+}p_{-}s_{+}s_{-}r_{1}r_{2}-
(R+2iq)\bar{p}_{+}\bar{p}_{-}\bar{s}_{+}\bar{s}_{-}r_{3}r_{4}\right]\\
&-2R^{2}\left[(\sigma-iq)\bar{p}_{+}\bar{p}_{-}s_{+}s_{-}r_{1}r_{3}-
(\sigma+iq)p_{+}p_{-}\bar{s}_{+}\bar{s}_{-}r_{2}r_{4}\right] \\
& - 2iq(R^{2}-4\sigma^{2}){\rm Re}\left(p_{+}\bar{p}_{-}s_{+}\bar{s}_{-}\right)(r_{1}r_{4}+r_{2}r_{3})\\
&-i\sigma R \{(R-2\sigma) {\rm Im}(p_{+}\bar{p}_{-}) \left[\bar{ \kappa}_{+}s_{+}s_{-}r_{1}+\kappa_{+}\bar{s}_{+}\bar{s}_{-}r_{4}\right]\\
&+(R+2\sigma) {\rm Im}(s_{+}\bar{s}_{-}) \left[\kappa_{-}p_{+}p_{-}r_{2}+\bar{\kappa}_{-}\bar{p}_{+}\bar{p}_{-}r_{3}\right] \},\\
p_{\pm}&:=2(M^{2}-q^{2})-(R \pm 2M)\sigma \pm M R +i [q(R-2\sigma)\pm \delta], \\
s_{\pm}&:=2(M^{2}-q^{2})+(R \mp 2M)\sigma \mp M R +i [q(R+2\sigma) \mp \delta],\\
\kappa_{o}&:=(R^{2}-4\sigma^{2})[(R^{2}-4M^{2})(M^{2}-\sigma^{2})+4q^{4}+4 M q\delta],\\
\kappa_{\pm}&:=R \pm 2(\sigma+2iq),
\label{3-parameters}  \end{split} \eea

\noi with
\bea \begin{split} r_{1,2}&=\sqrt{\rho^{2}+\left(z-R/2 \mp \sigma\right)^{2}}, \\
r_{3,4}&=\sqrt{\rho^{2}+\left(z+R/2 \mp \sigma\right)^{2}}. \end{split}\eea

In the above metric Eq.\ (\ref{3-parameters}) the condition $\sigma^{2}\geq 0$ describes BHs, while $\sigma^{2}< 0$ defines hyperextreme sources (relativistic disks). On the other hand, regarding the spacetime properties of the solution, it should be observed that the aforementioned Eq.\ (\ref{3-parameters}) contains a reflection-symmetric property \cite{Kordas}, since the change $z\rightarrow -z$ maintains invariant the metric functions $f$, $\omega$ and $\gamma$, while the Ernst potential on the symmetry axis given by
\bea \begin{split}
e(z)&=\frac{e_{+}}{e_{-}},\\
 e_{\pm}&=z^{2}\mp 2(M \pm iq)z + 2(M^{2}-q^{2})-R^{2}/4-\sigma^{2} \pm i \delta, \label{ernstaxis}\end{split}\eea

\vspace{-0.3cm}\noi satisfies the relation $e(z)\bar{e}(-z)=1$.

It is worth mentioning that Eq.\ (\ref{3-parameters}) is represented by only three parameters $\{R, M, q\}$, where the angular momentum $J$ enters explicitly into the solution by means of the real parameter $q$. It is computed from Eq.\ (\ref{ernstaxis}) via the Fodor-Hoenselaers-Perj\'es procedure \cite{FHP}; it reads
\be J=2Mq-\frac{\delta}{2}= \frac{M q [(R+2M)^{2}+4q^{2}]}{R(R+2M)+4q^{2}}, \label{momentum}\ee

\noi and leads us to a cubic equation given by
\be q^{3}-\frac{J}{M}q^{2}+ \frac{(R+2M)^{2}}{4}q- \frac{R(R+2M)J}{4M}=0, \label{theq}\ee

\noi whose explicit real solution is
\bea \begin{split}
q&=\frac{J}{3M}+a\left[\sqrt{b^{2}+a^{3}}-b\right]^{-1/3}-\left[\sqrt{b^{2}+a^{3}}-b\right]^{1/3}, \\
a&:=\frac{(R+2M)^{2}}{12}-\left(\frac{J}{3M}\right)^{2}, \\
b&:=\left(\frac{J}{3M}\right)\left[\frac{(R-M)(R+2M)}{4}+\left(\frac{J}{3M}\right)^{2}\right]. \label{theq} \end{split}\eea

Due to the fact that both BHs are identical, the event horizon for the upper BH is defined as a null hypersurface $H=\{-\sigma\leq z - \frac{R}{2}\leq \sigma,\,0 \leq \varphi \leq 2\pi,\, \rho\rightarrow 0\}$. Therefore, the Komar parameters \cite{Komar} are obtained through the Tomimatsu's formulae \cite{Tomimatsu}
\bea \begin{split}
M&=-\frac{1}{8\pi}\int_{H} \omega\, {\rm{Im}}({\cal{E}}_{z}) d\varphi dz, \\ J&=-\frac{1}{8\pi}\int_{H}\omega\, \left(1+\frac{1}{2}\omega\, {\rm{Im}}({\cal{E}}_{z}) \right) d\varphi dz.\label{Tomi}\end{split}\eea

\noi Replacing Eq.\ (\ref{3-parameters}) into Eq.\ (\ref{Tomi}), it can be demonstrated that $M$ and $J$ represent \emph{exactly} the mass and angular momentum, respectively, for each BH. So, there is no cast of doubt that the event horizon $\sigma$ given by Eq.\ (\ref{horizon}) is fully depicted by the physical parameters $\{R, M, J\}$. The total mass and total angular momentum of the system are $2M$ and $2J$ respectively.

By putting now our attention to the thermodynamical properties of the binary system, where each BH fulfills
the mass formula \cite{Smarr}
\be M=\frac{\kappa S}{4\pi} +2 \Omega J  =\sigma +2 \Omega J,  \label{Smarr}\ee

\noi where $\kappa$ is the surface gravity, $S$ the area of the horizon, and $\Omega$ the angular velocity. The surface gravity $\kappa$ and the angular velocity $\Omega$ are computed directly via the formulas \cite{Tomimatsu}
\be \kappa= \sqrt{- \Omega^{2} e^{-2\gamma^{H}}}, \qquad \Omega= 1/\omega^{H}, \ee

\noi being $\gamma^{H}$ and $\omega^{H}$ the respective values of the metric functions $\gamma$ and $\omega$ at the horizon. A straightforward calculation leads us to
\bea \begin{split}
\kappa&= \frac{\sigma (R+2\sigma)[R(R+2M)+4q^{2}]}{2M[(R+2M)^{2}+4q^{2}][(R+2M)(M+\sigma)-2q^{2}],}\\
\Omega&= \frac{J\{[R(R+2M)+4q^{2}]^{2}-4M^{2}(R^{2}-4M^{2}+4q^{2})\}}{2M^{2}(M+\sigma)[(R+2M)^{2}+4q^{2}]^{2}}.
\label{Horizonproperties} \end{split}\eea

In addition, $S$ is obtained from Eq.\ (\ref{Smarr}) with aid of Eq.\ (\ref{Horizonproperties}),
\bea \begin{split}
S&= \frac{4\pi M[(R+2M)^{2}+4q^{2}]}{R(R+2M)+4q^{2}}\\
&\times \left[R+2M-\frac{R^{2}-4M^{2}+4q^{2}}{R+2\sigma}\right].  \label{Areaofthehorizon} \end{split}\eea

Another physical property of this two-body configuration is the interaction force associated with the strut (conical singularity). It can be calculated by means of the formula \cite{Israel,Weinstein}
\bea \begin{split}
\mathcal{F}&=\frac{1}{4}(e^{-\gamma_{s}}-1)\\
&=\frac{M^{2}[(R+2M)^{2}-4q^{2}]}{[R^{2}-4M^{2}+4q^{2}][(R+2M)^{2}+4q^{2}]},
\label{force} \end{split}\eea

\noi with $\gamma_{s}$ as the metric function $\gamma$ evaluated on the region of the strut. At this point we observe already from Eqs.\ (\ref{horizon}) and (\ref{force}), the existence of a minimal distance value given by $R_{min}=2 \sqrt{M^{2}-q^{2}}$, on which both horizons are touching each other and the interaction force $\mathcal{F}\rightarrow \infty$. Moreover, such critical distance implies from Eqs.\ (\ref{momentum}) and (\ref{horizon}) that $q=J/2M$ and $\sigma= \sqrt{M^{2}-(J/2M)^{2}}$ respectively, and thus, the minimal interaction distance results to be
\be R_{min}=2\sigma \equiv 2 \sqrt{M^{2}-\left(\frac{J}{2M}\right)^{2}}, \label{minimal}\ee

\noi and thereby, at this particular distance arise the following limit values for $\kappa$, $S$, and $\Omega$,  given by
\bea \begin{split} \kappa&=\frac{\sigma}{4M(M+\sigma)}, \qquad S=16 \pi M(M+\sigma),\\  \Omega&=\frac{J}{8M^{2}(M+\sigma)}. \label{criticalvaluesI}\end{split}\eea

On the other hand, if $R \rightarrow \infty$ the interaction force vanishes ($\mathcal{F}\rightarrow 0$), and $q=J/M$ [see Eq.\ (\ref{momentum})]. For this case we recover from Eq.\ (\ref{horizon}) the expression of the horizon for one single Kerr BH
\be \sigma= \sqrt{M^{2}-\frac{J^{2}}{M^{2}}},\ee

\noi but now one obtains the following limit values for $\kappa$, $S$, and $\Omega$:
\bea \begin{split} \kappa&=\frac{\sigma}{2M(M+\sigma)}, \qquad S=8 \pi M(M+\sigma),\\  \Omega&=\frac{J}{2M^{2}(M+\sigma)}. \label{criticalvaluesII}\end{split}\eea

\begin{figure}[ht]
\centering
\includegraphics[width=8.5cm,height=5.0cm]{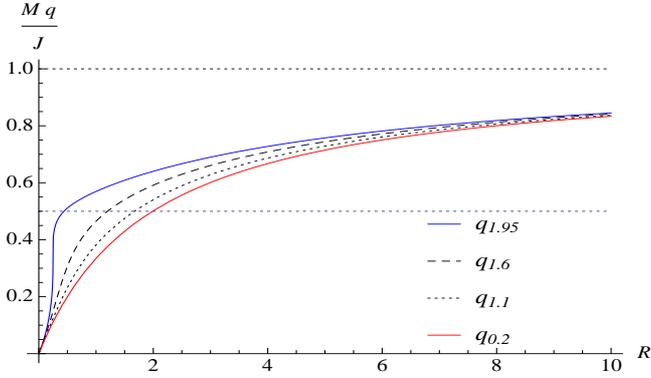}\\
\caption{Typical shapes of $q$ in the non-extreme case, for $M=1$, and different angular momentum values indicated by the subindex. The minimum value $q_{min}=J/2M$ arises at the distance $R_{min}=2 \sqrt{M^{2}-(J/2M)^{2}}$, while the maximum value $q_{max}=J/M$, if $R \rightarrow \infty$.}
\label{qnonextreme}\end{figure}

\begin{figure}[ht]
\centering
\includegraphics[width=8.5cm,height=5.0cm]{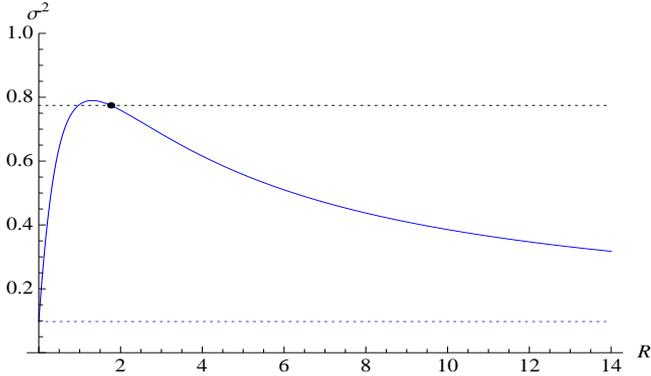}\\
\caption{For fixed mass $M=1$ and angular momentum value $J=0.95$, the condition $\sigma^{2}\geq0$ is ensured for coordinate distance values running within the interval $1.760\leq R< \infty$. The allowed maximum value for $\sigma^{2}=0.774$, while the minimum value is $\sigma^{2}=0.098$; both values are indicated by asymptotes.}
\label{sigmacuadrada1}\end{figure}

\begin{figure}[ht]
\centering
\includegraphics[width=8.5cm,height=5.0cm]{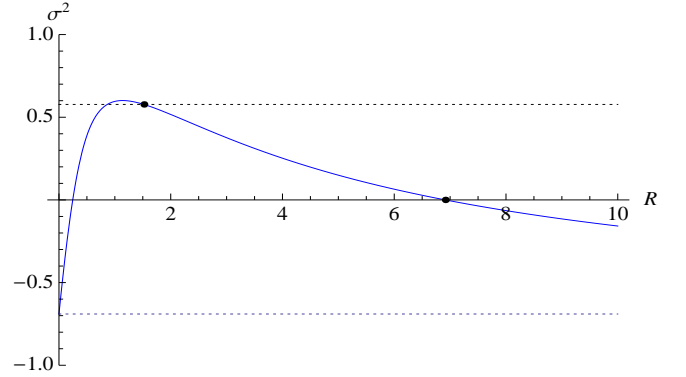}\\
\caption{For angular momentum value $J=1.3$ and $M=1$, the function $\sigma^{2}$ oscillates between the values $0.774$ and $-0.690$, respectively. Nevertheless, the condition $\sigma^{2}\geq0$ is ensured into the interval $1.520\leq R< 6.914$. Conforming $J\rightarrow 2M^{2}$ the interval of $R$ shrinks to zero, since it corresponds to the extreme limit case.}
\label{sigmacuadrada2}\end{figure}

Continuing with the analysis, several curves depicting the parameter $q$ are plotted in Fig.\ \ref{qnonextreme} for different values of the angular momentum. In fact $q$ grows monotonically taking real values within the range $J/2M\leq q \leq J/M$. On one hand, if we fixed the angular momentum value in the domain $0\leq J\leq M^{2}$, the condition $\sigma^{2}\geq0$ is ensured for all the coordinate distance values within the interval $R_{min}\leq R < \infty$. On the other hand, inside the values $M^{2}<J\leq 2M^{2}$ the domain of $R$ is shortened between the value $R_{min}$ and the one at which the extremality condition occurs, i.e., $\sigma=0$. These properties can be noticed in Figs.\ \ref{sigmacuadrada1} and \ref{sigmacuadrada2}.

\vspace{-0.3cm}
\vspace{-0.3cm}

It should be pointed out, that the minimal distance value defined by Eq.\ (\ref{minimal}), and therefore, the subsequent analysis leading to Eq.\ (\ref{criticalvaluesI}), was first derived \emph{numerically} by Costa \emph{et al.} \cite{Costa}. Such value was named as the \emph{merging limit}. However, throughout their work, those authors referred to an unknown function $\tilde{f}\equiv \tilde{f}(R,M,J)$, which was the key to describe the thermodynamical aspects of this corotating system. Because we have been working within the framework of an exact solution we know already the explicit formula for such function, namely
\be \tilde{f}=\frac{[R(R+2M)+4q^{2}]^{2}-4M^{2}(R^{2}-4M^{2}+4q^{2})}{[(R+2M)^{2}+4q^{2}]^{2}}, \ee

\noi where $q$ is given explicitly in Eq.\ (\ref{theq}). Thereby, the mystery of \cite{Costa} on the explicit form of the function  $\tilde{f}$, has been revealed in this work. Its typical shape is shown below in Fig.\ \ref{thef}.
\vspace{-0.3cm}
\vspace{-0.3cm}
\begin{figure}[ht]
\centering
\includegraphics[width=8.5cm,height=5.0cm]{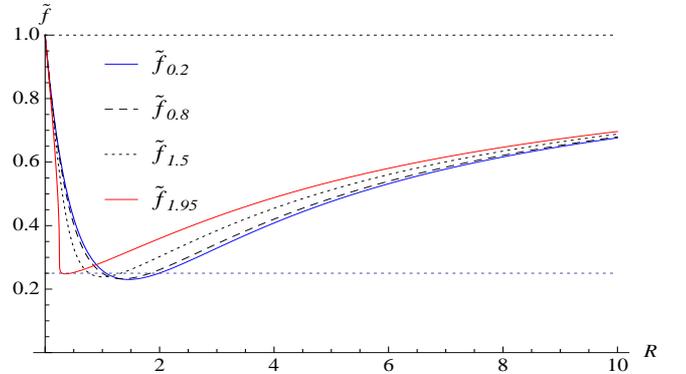}\\
\caption{The function $\tilde{f}$ is drawn for different values of the angular momentum, indicated below with a subindex. The starting value for $\tilde{f}=1/4$ emerges at $R_{min}$, while the maximum value $\tilde{f}=1$ is reached asymptotically if $R \rightarrow\infty$, in agreement with Ref.\ \cite{Costa}. }
\label{thef}\end{figure}

To conclude this section, the conical singularity in between sources can be removed by establishing the condition $\mathcal{F}=0$. In this situation $q=R/2+M>0$, thereby Eq.\ (\ref{horizon}) results to be
\be  \sigma^{2}=-\frac{R^{2} (R^{2}+6 M R+7M^{2})}{4(R+M)^{2}} \leq 0, \label{thedisk} \ee

\noi where now the sources represent two disks lying on the equatorial plane for $M>0$ (see Fig.\ \ref{DKidentical}). However, the notion of treating with subextreme sources is recovered from Eq.\ (\ref{thedisk}) only whether the individual mass turns out to be negative; i.e., $M<0$, therefore, naked singularities appear as ring singularities off the axis \cite{Hennig,DH}. Additionally, at this particular value for $q$, the individual angular momentum arises immediately from Eq.\ (\ref{momentum}); it reads
\be J= \frac{M(R+2M)^{2}}{2(R+M)} \label{momentumeq}.\ee

\vspace{-0.3cm}\section{Extreme limit case: the Kinnersley-Chitre subfamily}
\vspace{-0.2cm}The well-known Kinnersley-Chitre (KCH) $5$-parametric exact solution \cite{KCH} represents the extreme limit of the DKN vacuum solution of Kramer and Neugebauer \cite{KramerNeugebauer}. It was introduced in terms of the real parameters $p_{o}$, $q_{o}$, $\gamma_{o}$, $\alpha_{o}$, $\beta_{o}$, with the first three parameters satisfying the conditions
\be p_{o}^{2}+ q_{o}^{2}=1, \qquad |e^{-i\gamma_{o}}|=1,  \ee

\noi where is used the subscript ``$o$'' to avoid any confusion among $q_{o}$ and $q$ as well as other variables that will be used in what follows in this paper. Then, in order to develop the extreme limit case, one requires the extremality condition $\sigma=0$ and a careful application of l'H\^{o}pital's rule in the full metric Eq.\ (\ref{3-parameters}). This task is quite complicated to reach from a technical point of view, since the metric function $\omega$ cannot be expressed in a simple manner. Fortunately for us, after following Perj\'es' ideas \cite{Perjes} on the factor structure of the well-known Tomimatsu-Sato spacetimes \cite{TS}, the full metric of the extreme solution can be depicted by four basic polynomial $\rho_{o}$, $\sigma_{o}$, $\pi_{o}$, and $\tau_{o}$, leading us to
\bea \begin{split} {\cal{E}}&=\frac{A-B}{A+B},\quad f=\frac{D}{N}, \quad
\omega=\frac{\alpha(y^{2}-1)W}{D},\\
e^{2\gamma}&=\frac{D}{\alpha^{8}(x^{2}-y^{2})^{4}}, \\
A&=\alpha^{2}\left[(\alpha^{2}-\Delta)(x^{2}-y^{2})^{2}+\Delta(x^{4}-1)\right]\\
&+ (q^{4}+\alpha^{2}M^{2}-M^{4}-2 M q \delta_{o})(1-y^{4}) + 2i \alpha x y\\
&\times \left\{2[q(\Delta-\alpha^{2})+M \delta_{o}](y^{2}-1)-\alpha^{2}q(x^{2}-y^{2}) \right\},\\
B&=2\alpha x \left\{\alpha^{2}M (x^{2}-y^{2})-[M\Delta+q \delta_{o}](1-y^{2})\right\}\\
&-2i y\{[M q(2\Delta-\alpha^{2})+(M^{2}+q^{2})\delta_{o}](1-y^{2})\\
& + \alpha^{2}\delta_{o}(x^{2}-y^{2}) \}, \\
D&= \rho_{o}^{2}+(x^{2}-1)(y^{2}-1)\sigma_{o}^{2},\\
N&= D+ \rho_{o} \pi_{o}-(1-y^{2})\sigma_{o} \tau_{o}, \\
W&=(x^{2}-1)\sigma_{o}\pi_{o}-\rho_{o} \tau_{o},\\
\rho_{o}&=\alpha^{2}[(\alpha^{2}-\Delta)(x^{2}-y^{2})^{2}+\Delta(x^{2}-1)^{2}]\\
&- (q^{4}+\alpha^{2}M^{2}-M^{4}-2 M q \delta_{o})(y^{2}-1)^{2},\\
\sigma_{o}&=2\alpha\left\{\alpha^{2}q(x^{2}-y^{2})+2[q(\alpha^{2}-\Delta)-M\delta_{o}]y^{2}\right\},\\
\pi_{o}&=4 \{ \alpha x [M(\alpha x+M)^{2}+q \delta_{o}(1+y^{2})-M q^{2}]\\
&-(\alpha^{2}-\Delta)[\alpha M x+ 2\Delta]y^{2}\},\\
\tau_{o}&=(4/\alpha)\{ ( 2Mq^{2}\delta_{o} + q(M^{4}-q^{4}-\alpha^{2}q^{2})\\
&+\alpha[M q(2\Delta-\alpha^{2})+(M^{2}+q^{2})\delta_{o}]x )(1-y^{2})\\
& +\alpha^{2} \delta_{o} (2M+\alpha x)(x^{2}-y^{2})\},\\
\delta_{o}&:= \sqrt{\Delta(\Delta-\alpha^{2})}, \quad \Delta:=M^{2}-q^{2},\quad \alpha:=\frac{R}{2},
\label{extreme}\end{split}\eea

\noi where the aforementioned solution Eq.\ (\ref{extreme}) is written in prolate spheroidal coordinates $(x,y)$ defined as
\bea \begin{split} x=\frac{r_{+}+r_{-}}{2\alpha}, \quad y=\frac{r_{+}-r_{-}}{2\alpha}, \quad
r_{\pm}=\sqrt{\rho^{2} +(z\pm\alpha)^{2}}, \label{prolates}\end{split}\eea

By setting $\sigma=0$, Eq.\ (\ref{extreme}) is characterized by only two parameters, where the angular momentum of Eq.\ (\ref{momentum}) is explicitly defined in terms of the mass and coordinate distance as follows:
\bea \begin{split}
J&= \frac{M q [(\alpha+M)^{2}+q^{2}]}{\alpha(\alpha+M)+q^{2}}, \\
3q^{2}&=  \left[b_{o}+ \sqrt{b_{o}^{2}-a_{o}^{3}}\right]^{1/3}+a_{o} \left[b_{o}+ \sqrt{b_{o}^{2}-a_{o}^{3}}\right]^{-1/3}\\
&-2\left(\alpha^{2}-M^{2}+\alpha M\right),\\
a_{o}&:=(\alpha^{2}-M^{2})(\alpha^{2}-M^{2}+2\alpha M)+4\alpha^{2}M^{2},\\
b_{o}&:=(\alpha^{2}-M^{2}+\alpha M)^{3}+ \left(\frac{9}{2}\right)\alpha^{2}M^{2}(\alpha^{2}+4\alpha M+5M^{2}),
\label{momentumextreme}
\end{split}\eea

\begin{figure}[ht]
\centering
\includegraphics[width=8.5cm,height=5.0cm]{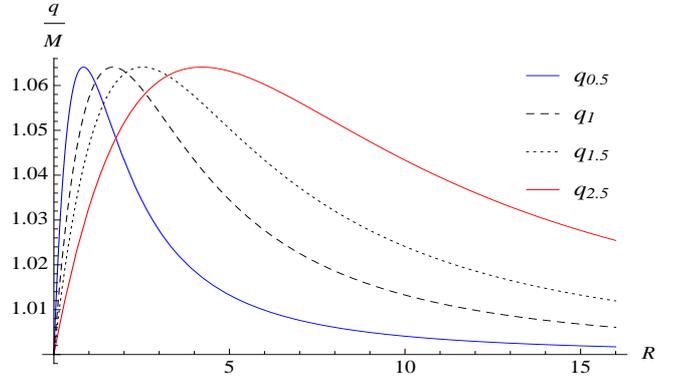}\\
\caption{Behavior for the parameter $q$ in the extreme case, taking different values in the mass $M$ denoted by the subscript. The maximum point is located approximately at $(1.6861M, 1.0641M)$.  }
\label{qextreme}\end{figure}
\vspace{-0.5cm}
\begin{figure}[ht]
\centering
\includegraphics[width=8.5cm,height=5.0cm]{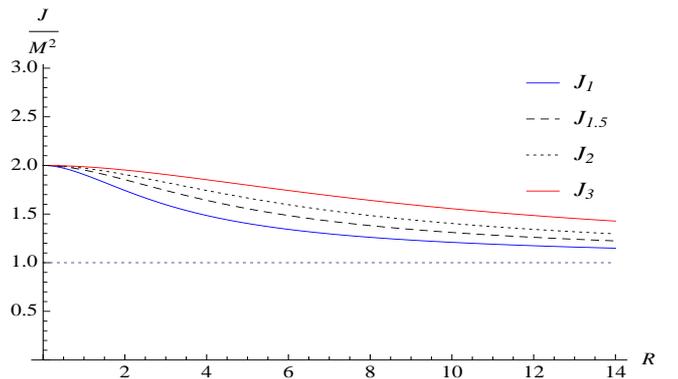}\\
\caption{The angular momentum for different mass values denoted in the subscript.  }
\label{Jext}\end{figure}

\noi but now the parameter $q$ is a function which starts and ends at the same value $M$, running from $R=0$ until $R \rightarrow \infty$. Besides, there exists a maximum value in $q\simeq 1.0641M$, given at $R \simeq 1.6861M$. All these characteristics can be observed in Fig.\ \ref{qextreme}. Moreover, if the extremality condition is achieved, the angular momentum for the extreme case given by Eq.\ (\ref{momentumextreme}) contains an aspect shown like in Fig.\ \ref{Jext}. Such behavior of the angular momentum in the extreme case, was provided first in Ref.\ \cite{Costa} by using numerical methods, since those authors never derived an explicit formula for the relation $J/M^{2}\equiv \tilde{f}(R,M)^{-1/2}$ in terms of mass $M$ and coordinate distance $R$.

Lastly, our metric given by Eq.\ (\ref{extreme}) can be identified as a $2$-parametric KCH subfamily member, after doing the following change in the real parameters of formula (4.30) in Ref.\ \cite{KCH}:
\bea \begin{split}
\alpha_{o}&=\gamma_{o}=0, \qquad \beta_{o}=\frac{\alpha [q(\alpha^{2}-M^{2}+q^{2})-M\delta_{o}]}{M^{4}-q^{4}-\alpha^{2}q^{2}+2Mq\delta_{o}},\\
p_{o}^{2}&=\frac{\alpha^{2} (M^{2}-q^{2})}{M^{4}-q^{4}-\alpha^{2}q^{2}+2Mq\delta_{o}}= 1-q_{o}^{2},\\
\end{split}\eea

\noi where the physical parameters $\{R,M\}$ enter explicitly inside the solution, via the parameter $q$ given in Eq.\ (\ref{momentumextreme}).

\vspace{-0.3cm}\section{Concluding remarks}
\vspace{-0.2cm}We have succeeded in working out the full metric for a two-body system of identical corotating BHs apart by a massless strut, as a $3$-parametric asymptotically flat exact solution, where the horizon half-length parameter $\sigma$ as well as all the thermodynamical features contained into the Smarr mass formula \cite{Smarr} are explicitly expressed in terms of Komar physical parameters \cite{Komar}. All limits provided numerically in Ref.\ \cite{Costa} are obtained analytically in our paper. Furthermore, since the full metric was constructed by means of a suitable parametrization, it motivates us for searching new physical models on more sophisticated configurations including the electromagnetic field, not only for identical cases but also for unequal constituents, like the one performed in Ref. \cite{RICM} for unequal counterrotating Kerr-Newman sources. On the other hand, regarding the coalescence process among both Kerr sources, in the absence of a supporting strut, relativistic disks emerge in the binary model if the mass of each BH satisfies the condition $M>0$. Nevertheless, we observe from Eq.\ (\ref{thedisk}) that if $R=0$, the merging process converts the two disks into one extreme source; hence, there appears a single extreme Kerr BH of mass $M_{o}=2M$ and total angular momentum $J_{o}=2J$. For such a situation, the total angular momentum and the total mass of the new single BH satisfies the well-known relation for extreme BHs, i.e., $J_{o}=M_{o}^{2}$. The last formula is in agreement with the fact that the individual angular momentum satisfies the relation $J=2M^{2}$, after establishing $R=0$ in Eq.\ (\ref{momentumeq}). \footnote{The possibility that the individual black holes may violate the Kerr bound by means of the inequality $|J|> M^{2}$,
was first pointed out by Herdeiro \emph{et al.}, in identical counterrotating systems \cite{Herdeiro}.} Because our solution Eq.\ (\ref{3-parameters}) might represent relativistic disks under the change $\sigma \rightarrow i \sigma$, it would be interesting to deepen more in this subject in future works following Bardeen and Wagoner ideas \cite{Bardeen}.

Owing that our binary model contains a physical parametrization and takes into account the coalescence process which forms a single regular black hole without a conical singularity, we strongly believe that it can be considered as a first step to analyze geodesics around  binary BH systems and future researches regarding gravitational waves (GW) \cite{LIGO}, since the quasinormal modes of GW can be also studied from the physical point of view of free oscillations of unstable circular null geodesics in the geometric-optics (eikonal) limit \cite{Mashhoon, Sod}.

To conclude, after knowing our current research, in a recent preprint \cite{MR} was correctly pointed out a further simplification of our metric Eq.\ (\ref{3-parameters}), which is written down in a more compact form. To clarify this point, we note from Eq.\ (\ref{3-parameters}) that terms $p_{\pm}$ and $s_{\pm}$ can be arranged in such a way that

\be a_{1}:=\frac{s_{+}}{\bar{s}_{+}}, \quad a_{2}:=\frac{p_{-}}{\bar{p}_{-}}, \quad a_{3}:=\frac{\bar{p}_{+}}{p_{+}}, \quad a_{4}:=\frac{\bar{s}_{-}}{s_{-}}, \quad |a_{j}|\equiv1, \ee

\noi and because $a_{j}$ satisfy the relations $a_{1}=-\bar{a}_{4}$,\, $a_{2}=-\bar{a}_{3}$, the metric Eq.\ (\ref{3-parameters}) reduces considerably its aspect as follows:
\bea \begin{split}
{\cal{E}}&=\frac{\Lambda+\Gamma}{\Lambda-\Gamma},\quad
f=\frac{\Lambda \bar{\Lambda}-\Gamma \bar{\Gamma}}{(\Lambda-\Gamma)(\bar{\Lambda}-\bar{\Gamma})}, \\ \omega&=4q +\frac{2{\rm{Im}}\left[(\Lambda-\Gamma)(z\bar{\Gamma}+\bar{\mathcal{G}})\right]}{\Lambda \bar{\Lambda}-\Gamma \bar{\Gamma}},\quad e^{2\gamma}=\frac{\Lambda \bar{\Lambda}-\Gamma \bar{\Gamma}}{\kappa_{o}^{2} r_{1}r_{2}r_{3}r_{4}},\\
\Lambda&=R^{2}(\mathfrak{r}_{1}-\mathfrak{r}_{2})(\mathfrak{r}_{3}-\mathfrak{r}_{4})
-4\sigma^{2}(\mathfrak{r}_{1}-\mathfrak{r}_{3})(\mathfrak{r}_{2}-\mathfrak{r}_{4}),\\
\Gamma&=2\sigma R\left[(R-2\sigma)(\mathfrak{r}_{1}-\mathfrak{r}_{4})-(R+2\sigma)
(\mathfrak{r}_{2}-\mathfrak{r}_{3})\right],\\
\mathcal{G}&= 2\sigma R \left[ R(\mathfrak{r}_{1}\mathfrak{r}_{3}-\mathfrak{r}_{2}\mathfrak{r}_{4})
-2\sigma(\mathfrak{r}_{1}\mathfrak{r}_{2}-\mathfrak{r}_{3}\mathfrak{r}_{4})\right]\\
&+\sigma R (R^{2}-4\sigma^{2}) (\mathfrak{r}_{1}-\mathfrak{r}_{2}-\mathfrak{r}_{3}+\mathfrak{r}_{4}), \qquad \mathfrak{r}_{j}:=a_{j}r_{j},\\
\kappa_{o}&:= \frac{4\sigma^{2}R^{2}(R^{2}-4\sigma^{2})}{(R^{2}+4q^{2})(\sigma^{2}+q^{2})-4M^{4}-4M q\delta}.
\label{new3-parameters}  \end{split} \eea

Nonetheless the solving of axis conditions is the main challenge to study dynamical and physical properties of these configurations. Once we know a specific axis data, the SM provides the solution in the whole spacetime. We expect to develop in a future some extensions of the present model including the electromagnetic field.

\vspace{-0.3cm}\section*{Acknowledgements}
\vspace{-0.2cm}This work was supported by the SNI program from CONACYT, M\'exico, grant No. 56244 and by Fondo Santander para docentes from Universidad Aut\'onoma de Ciudad Ju\'arez (UACJ). ICM also acknowledges the financial support of PROMEP.

%\section*{References}

\end{document}